# Voltage-Controlled Magnon Transistor via Tunning Interfacial Exchange Coupling

Y. Z. Wang[#], T. Y. Zhang[#], J. Dong, P. Chen, C. H. Wan*, G. Q. Yu, X. F. Han*

[1]Beijing National Laboratory for Condensed Matter Physics, Institute of Physics, University of Chinese Academy of Sciences, Chinese Academy of Sciences, Beijing 100190, China

[2]Center of Materials Science and Optoelectronics Engineering, University of Chinese Academy of Sciences, Beijing 100049, China

[3]Songshan Lake Materials Laboratory, Dongguan, Guangdong 523808, China

*Email: xfhan@iphy.ac.cn; wancaihua@iphy.ac.cn

Abstract: Magnon transistors that can effectively regulate magnon transport by an electric field are desired for magnonics which aims to provide a Joule-heating free alternative to the conventional electronics owing to the electric neutrality of magnons (the key carriers of spin-angular momenta in the magnonics). However, also due to their electric neutrality, magnons have no access to directly interact with an electric field and it is thus difficult to manipulate magnon transport by voltages straightforwardly. Here, we demonstrated a gate voltage ($V_g$) applied on a nonmagnetic metal/magnetic insulator (NM/MI) interface that bended the energy band of the MI and then modulated the possibility for conduction electrons in the NM to tunnel into the MI can consequently enhance or weaken the spin-magnon conversion efficiency at the interface. A voltage-controlled magnon transistor based on the magnon-mediated electric current drag (MECD) effect in a Pt/$Y_3Fe_5O_{12}$ (YIG)/Pt sandwich was then experimentally realized with $V_g$ modulating the magnitude of the MECD signal. The obtained efficiency (the change ratio between the MECD voltage at $\pm V_g$) reached 10%/(MV/cm) at 300 K. This prototype of magnon transistor offers an effective scheme to control magnon transport by a gate voltage.

Magnons as the collective excitation of a magnetically ordered lattice possess both spin-angular momenta and phases but no charges [1], born an ideal information carrier for the Joule-heating-free electronics [2,3]. In order to efficiently manipulate magnon transport, magnon transistors, as an elementary brick for magnonics, are long-desired. Despite of great achievements in efficiently exciting [4-10], propagating [11-13] and detecting [14-18] magnons, the electric neutrality of magnons sets a high level of difficulty in controlling magnon transport by electric fields.

Several magnon gating methods have been realized. The YIG/Au/YIG/Pt magnon valves [19] and YIG/NiO/YIG/Pt magnon junctions [20-22] are gated by *an external magnetic field* (*H*). Large (small) spin-Seebeck voltage was output by setting the two YIG layers into the parallel (antiparallel) state by *H* currently, though spin-orbit torques (SOT) are potentially used to gate the magnon valves/junctions in the future [23]. Another magnon-spin valve YIG/CoO/Co was also *H*-gateable [24]. Its spin pumping voltage depends on the *H*-controlled parallel/antiparallel states between YIG and Co. Another method is *gating current*. In the magnon transistor consisting of three Pt stripes on top of a YIG film [25], a charge current in the leftmost Pt stripe excites a magnon current in YIG via (i) the spin Hall effect (SHE) in Pt and (ii) the interfacial *s-d* coupling at the Pt/YIG interface. The as-induced magnon current diffuses toward the rightmost Pt stripe where the inverse process occurs, resulting in a detectable voltage. This phenomenon, featured by the nonlocal electric induction across a MI with the help of magnons, is named as the magnon-mediated electric current drag (MECD) effect [4,6]. A *gating current* flowing in the middle Pt strip changes magnon density of YIG in the gate region and consequently modifies the MECD efficiency. Gate voltage is advantageous in energy consumption. However, due to no direct coupling of magnons with any electric fields, magnon transistors inherently controlled by $V_g$ are still missing.

Inspired by the model of Chen *et al.* [26,27], we realize the spin mixing conductance ($G_{\uparrow\downarrow}$) at a NM/MI interface relies sensitively on the interfacial *s-d* exchange coupling. Here, we proposed a voltage-gated magnon transistor (Fig.1(a)) where $V_g$ across the NM/MI interface tilts downward (upward) the energy band of the MI (Fig.1(b)), decreases (increases) the probability of electrons penetrating into the MI (Fig.1(c)), thus weakens (strengthens) the spin-magnon conversion efficiency at the interface and consequently changes the magnon excitation efficiency in the magnon transistor.

We extended the model by including the $V_g$-induced band bending of MI via the Hamiltonian

$$H_{\mathrm{MI}} = p^2/2m + V_0 + \Gamma \mathbf{S} \cdot \boldsymbol{\sigma} + ezV_g/t \tag{1}$$

, where $V_0$ is the energy barrier at the interface; $\Gamma \mathbf{S} \cdot \boldsymbol{\sigma}$ describes the *s-d* coupling of electron spins $\boldsymbol{\sigma}$ in NM with localized moments **S** in MI; $ezV_g/t$ describes the conduction band bending by $V_g$ and *t* is the MI thickness (calculation details in Supplementary Materials). The predicted $V_0$ and electric field $E$ ($E = V_g/t$) dependence of the real part of $G_{\uparrow\downarrow}$ ($G_r$) was plotted in Fig.1(d) (taking Fermi energy of NM $\varepsilon = 5$ eV and s-d coupling strength $\Gamma=0.5$ eV for the Pt/YIG interface [4]). The typical $G_r$-$E$ curves at three $V_0$ values (Fig.1(e)) suggest the positive $V_g$ can efficiently increase $G_r$ and vice versa. The spin-magnon convertance at the NM/MI interface is proportional to $G_r$ [28] and thus the magnon current excited in MI can be modified by $V_g$ as experimentally shown below.

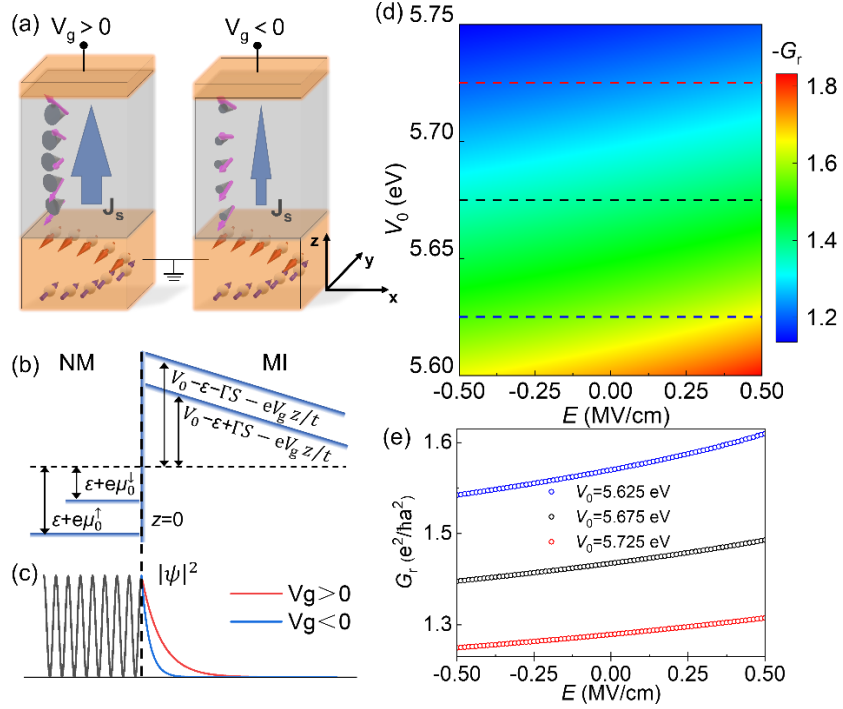

Fig.1. Mechanism of voltage-gated magnon transistor. (a) Schematics of the voltage-gated magnon transistor. A spin current is generated by the spin Hall effect (SHE) in the bottom (B)-NM, which induces imbalanced spin accumulation ($\mu_s$) at the B-NM/MI interface. Due to the s-d exchange coupling at the interface, $\mu_s$ relaxes by annihilating (generating) magnons in MI as $\mu_s$ has parallel (antiparallel) polarization to the magnetization of MI. The excited magnon current was thus manipulated by $V_g$: positive (negative) $V_g$ increases (decreases) its magnitude. (b) Schematics of potential profile near the B-NM/MI interface under positive $V_g$. (c) Schematics of probability $|\varphi|^2$ at the B-NM/MI interface under positive and negative $V_g$. (d) The predicted $V_0$ and $E$ dependence of $G_r$ (the color scale bar in units of $e^2/\hbar a^2$). (e) The $E$-dependence of $G_r$ under $V_0$=5.625, 5.675 and 5.725 eV extracted from Fig.1(d).

The $V_g$-controlled magnon transistor was then experimentally achieved in a Pt(10)/YIG(80)/Pt(5 nm) sandwich (details in Method and Supplementary Material) where $V_g$ across the YIG was able to tune the MECD effect. The measured voltage $V$ along the top (T)-Pt electrode follows the coming 3 characteristics: (1) the angular dependence of $V = V_{\text{drag}} \cos 2\theta$ ($\theta$ the angle between spin polarization $\sigma$ and magnetization $\mathbf{M}$, Fig.S6(a)), (2) the linear dependence of $V_{\text{drag}}$ on the input current ($I_{\text{in}}$) along the B-Pt electrode (Fig.S6(b,c)) and (3) the $T^{5/2}$ temperature-dependence (Fig.S6(d)), all coinciding with Ref.6&7 [7,8]. These features confirmed the MECD nature of the measured voltage. The insulating property of YIG was also checked by $I_{\text{leak}} - V_{\text{g}}$ curves (Fig.S3(b)) with the leakage current $I_{\text{leak}}$. $I_{\text{leak}}$ was independent on $H$, assuring the irrelevance of the observed $H$-dependent $V$ with $I_{\text{leak}}$ (Fig.S4).

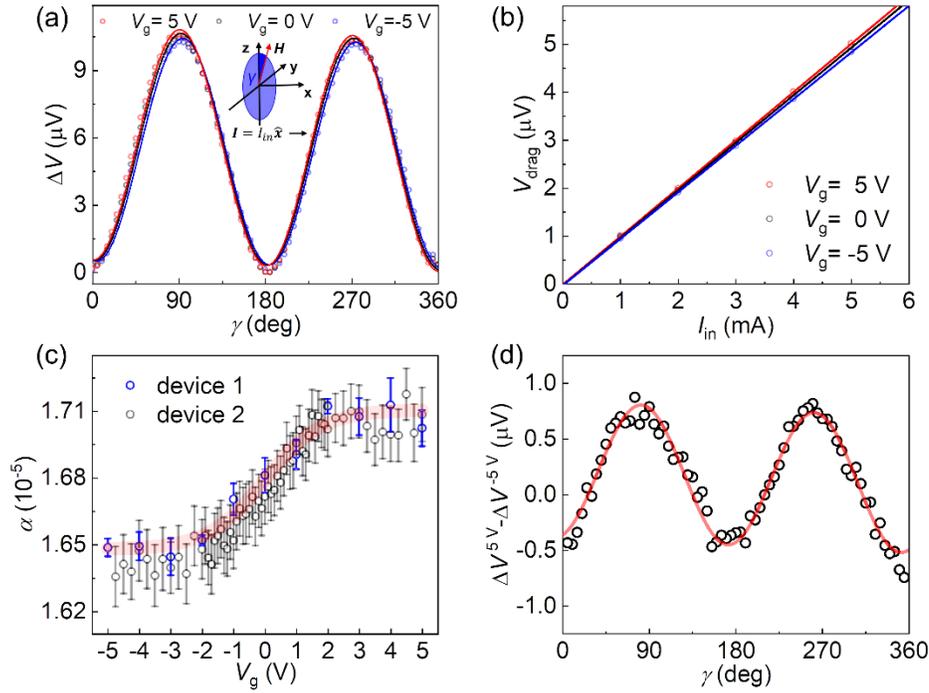

Fig.2. Voltage-controlled MECD effect. (a) The $\gamma$-dependence of $\Delta V$ with $\mathbf{H}$ rotated in the yoz plane and $I_{\text{in}} = 5$ mA under $V_{\text{g}} = -5, 0$ and $+5$ V. The open circles (solid lines) are the experimental data (fitted curves by $\Delta V = V_{\text{drag}} \cos 2\gamma$). (b) The $I_{\text{in}}$-dependences of $V_{\text{drag}}$ under different $V_{\text{g}}$ and their linear fittings. (c) The $V_{\text{g}}$-dependence of magnon drag parameter $\alpha$. Error bars for Device 1 and 2 are from the standard deviations of the linear fittings of the $V_{\text{drag}} - I_{\text{in}}$ relation and the $\Delta V = V_{\text{drag}} \cos 2\gamma$ fittings, respectively. The red line is the hyperbolic tangent fitting of the $\alpha$-$V_{\text{g}}$ curve. (d) The $\gamma$-dependence of the difference in $\Delta V$ between $V_{\text{g}} = \pm 5$ V.

The $V_{\text{g}}$-controllability of the MECD effect is clearly shown in Fig.2. The MECD magnitude was noticeably enhanced (weakened) under $V_{\text{g}} = +5$ V $(-5$ V) (Fig.2(a)), which was further confirmed by the slope change of the $V_{\text{drag}} - I_{\text{in}}$ curves (Fig.2(b)). The magnon drag parameter $\alpha$ was then calculated by $\frac{V_{\text{drag}}}{R_{\text{T-Pt}}} = \alpha I_{in}$. The $V_g$-dependence of the extracted $\alpha$ (see method) (Fig.2(c)) showed a clear change as $V_g = [-2\text{ V}, +2\text{ V}]$ and nearly saturated beyond the region. The maximum $\alpha$ tunability by $V_g$ $\left(\frac{\alpha(V_g>+2V)-\alpha(V_g<-2V)}{\alpha(V_g<-2V)}\right)$ reached ~ 5% with $\alpha(V_g > +2V) \sim 1.71 \times 10^{-5}$ and $\alpha(V_g < -2V) \sim 1.63 \times 10^{-5}$. The $\alpha$-controllability by $V_g$ was also repeated in another Device 2. In order to trace the trend of the $V_g$-induced change in $\alpha$, we fitted the $\alpha$-$V_g$ curve by a hyperbolic tangent function $\alpha = a + b\tanh(cV_g)$ as shown by the red line in Fig.2(c). Note that this fitting only mathematically impacts with $|b/a|$ and $c$ reflecting the magnitude and saturation speed of the $V_g$-tunability, respectively. Here, for the $\alpha$-$V_g$ curve $|b/a|=0.019$ and $c=-0.55$ V$^{-1}$.

In the following, we reveal the origin of the $V_g$-tunability over the MECD effect. First, the $V_g$-dependence of the MECD effect cannot be caused by any magnon coupling possibilities with the leakage current since $I_{\text{leak}}$ increased divergently with the increase in $|V_g|$ but $\alpha$ nearly saturated above $\pm 2$ V. Second, the resistance of T-Pt directly changed by $V_g$ was negligibly small (<0.008%, Fig.S8), also impossible to cause such significant change ~5% in the MCD signal. Third, though negligibly small in garnets [29-31], the interfacial Dzyaloshinsky-Moriya interaction (DMI) may introduce an additional magnon-drift velocity $\mathbf{v}_{\text{DMI}} = \hat{\mathbf{z}} \times \hat{\mathbf{m}}\frac{2\gamma}{M_s}D$ to influence magnon transport with $\hat{\mathbf{z}}$ the interfacial normal, $\hat{\mathbf{m}}$ ($M_s$) the magnetization direction (saturated magnetization), $\gamma$ the gyromagnetic ratio and $D$ a $V_g$-changeable parameter quantifying the DMI [32-34]. However, this DMI mechanism, if any, would bring about a 360° period in the yoz rotation owing to the $\hat{\mathbf{m}}$-dependence of $\mathbf{v}_{\text{DMI}}$. In stark contrast, the $\Delta V^{+5\,V} - \Delta V^{-5\,V}$ vs $\gamma$ curve (Fig.2d) shows a $\cos2\gamma$ symmetry (180° period), thus ruling out the DMI origin of the $V_g$ controllability.

To be more specific, the MECD effect can be explicitly expressed as below [4,6]:

$$\mathbf{j}_e^{\text{T-Pt}} \propto \theta_{\text{SH}}^{\text{top}}\theta_{\text{SH}}^{\text{bottom}} G_S^{s-m} G_S^{m-s} \boldsymbol{\sigma} \times (\mathbf{M} \times \mathbf{j}_e^{\text{B-Pt}}) \qquad (2)$$

here, $\mathbf{j}_e^{\text{T-Pt}}$ ($\mathbf{j}_e^{\text{B-Pt}}$) is the induced (input) charge current density along the T-Pt (B-Pt) electrode, $\theta_{\text{SH}}^{\text{top(bottom)}}$ is the spin Hall angle of the top (bottom) Pt electrode, $G_S^{s-m}$ ($G_S^{m-s}$) is the effective spin-magnon (magnon-spin) convertance at the B-Pt/YIG (YIG/T-Pt) interface, $\boldsymbol{\sigma}$ is the spin polarization perpendicular to $\mathbf{j}_e^{\text{T-Pt}}$ and $\mathbf{M}$ is the YIG magnetization. Ruling out the above 3 reasons, the MECD voltage can still be potentially manipulated by $V_g$ in the following scenarios: (1) $V_g$-induced changes in the effective magnetization of YIG, (2) the spin Hall angles ($\theta_{SH}$) of Pt or (3) the spin-magnon conversion efficiency across the B-Pt/YIG or YIG/T-Pt interfaces. Hereafter, we experimentally check their possibilities one-by-one.

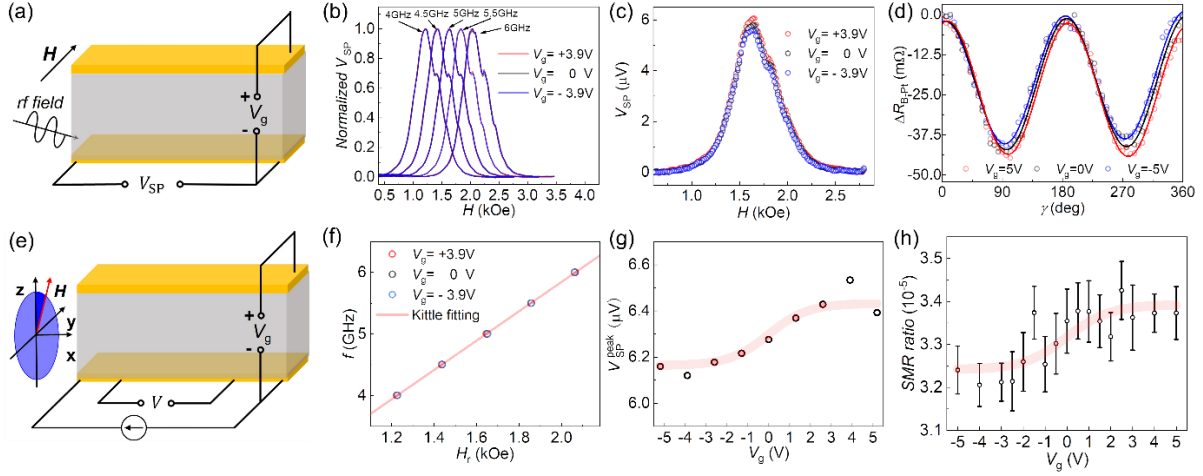

Fig.3. Schematics setups for (a) spin pumping measurement where the spin pumping voltage ($V_{SP}$) was picked up along the B-Pt stripe with **H** perpendicular to the stripe and $V_g$ applied across the sandwich and for (e) SMR measurement where the resistance change $\Delta R_B$ of the B-Pt stripe was measured with **H** rotated in the yoz plane. (b) The $H$-dependence of the normalized $V_{SP}(H)/V_{SP}^{max}$ under different $rf$ frequencies ($f$) and $V_g$=-3.9, 0 and +3.9 V. (c) The $H$-dependence of $V_{SP}$ at $f=5$ GHz and $V_g = -3.9, 0, +3.9$ V. (Error bars from standard deviation by fitting $V_{SP} - H$ curves with the Lorentzian function.) (d) The $\gamma$-dependences of the $\Delta R_B$ (open circles) and their $\Delta R_B = \Delta R_{SMR} \cos 2\gamma$ fittings. (f) The resonance field ($H_r$) dependence of $f$ under $V_g = -3.9, 0$ and $+3.9$ V (open circles) and their Kittle fittings. The $V_g$-dependence of (g) the peak value of $V_{SP} - H$ curve ($V_{SP}^{peak}$) under $f=5$ GHz and (h) the SMR ratio. (Error bars from standard deviation of the $\Delta R_B = \Delta R_{SMR} \cos 2\gamma$ fittings.) Red lines in Fig.3(c&d) are the hyperbolic tangent fitting of the $V_{SP}^{peak}$-$V_g$ and *SMR ratio*-$V_g$ curves, respectively.

To investigate the $V_g$-dependence of $M_s$, we conducted spin pumping experiments (experimental details in Method). The spin pumping voltage $V_{SP}$ picked up in the B-Pt electrode at various $V_g$ is exhibited in Fig.3(a). The $H$-dependences of a normalized $V_{SP}$ at different $f$ and $V_g$ show no noticeable changes (variation<0.3%) in the resonance field ($H_r$) (Fig.3(b)) and the overlapped Kittle fittings manifested no changes in the magnetization and anisotropy of YIG under $V_g$. Interestingly, the magnitude of $V_{SP}^{peak}$ changed by $V_g$ (Fig.3(c)). The tunability defined by $\frac{V_{SP}^{peak}(V_g=+3.9\,V) - V_{SP}^{peak}(V_g=-3.9\,V)}{V_{SP}^{peak}(V_g=-3.9\,V)}$ was also ~5 %. Moreover, the $V_{SP}^{peak}$-$V_g$ tendency seemed similar to the $V_{drag}$-$V_g$ relation, with $|b/a|=0.021$ and $c=-0.54$ V$^{-1}$ extracted from the hyperbolic tangent fitting. We also tested $V_{SP}$ along the T-Pt stripe, which had ideally identical $H_r$ but opposite polarity with the B-Pt stripe (Fig.S7(a)). However, $V_{SP}^{peak}$ was not changed by $V_g$ for the T-Pt detector (Fig.S7(c)). Since spin currents were both pumped out from the sandwiched YIG, the different $V_g$-controllability on $V_{SP}$ for the B-Pt and the T-Pt detectors strongly hinted an interfacial gating origin instead of any bulk YIG reasons.

The following $V_g$-dependent spin Hall magnetoresistance (SMR) effect also supported this interfacial claim. Since SMR originates from spin-transfer at interfaces and shunted by a thick Pt layer, we fabricated another Pt(4)/YIG(80)/Pt(5 nm) sandwich. Its $\Delta R_{B-Pt}$-$\gamma$ relation at various $V_g$ and the summarized $V_g$-dependence of the SMR ratio are shown in Fig.3(d,h). The similar coefficients of $|b/a|=0.022$ and $c=-0.51$ V$^{-1}$ were obtained from the hyperbolic tangent fitting (the red line in Fig.3(h)), illustrating the $V_g$-tunability on the SMR ratio also followed the similar trend as the $V_g$-dependence of $\alpha$ and $V_{SP}^{peak}$. The SMR effect in the T-Pt stripe was independent on $V_g$ (Fig.S7(b,d)).

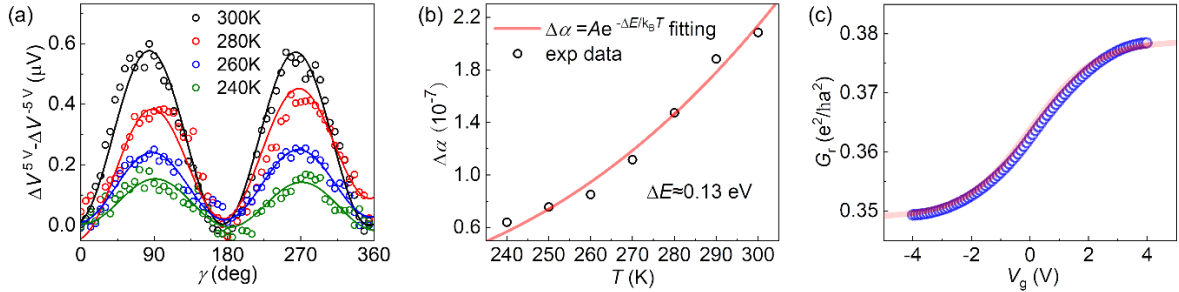

Fig.4. (a) The $\gamma$-dependence of $\Delta V(V_g=5V)-\Delta V(V_g=-5V)$ under different $T$. (b) The $T$-dependence of the difference in the magnon drag parameter $\Delta\alpha=\alpha(V_g=5\,V)-\alpha(V_g=-5\,V)$ between $V_g=\pm 5$ V. The solid lines are obtained by fitting data using $\Delta\alpha = Ae^{-\Delta E/k_B T}$. (c) The calculated $V_g$-dependence of $G_r$ by taking redistributed voltage on the contact resistance ($R_{contact}$) into consideration. The red line is the hyperbolic tangent fitting result.

After the above analysis we have narrowed possibility for the $V_g$-controlled MCD effect to (1) a $V_g$-changeable spin Hall angle in B-Pt or (2) a $V_g$-controllable spin-magnon conversion efficiency across the B-Pt/YIG interface. If the bulk spin Hall angle was modulated by $V_g$, we would not expect a substantial difference between the B-Pt and T-Pt stripes since they were both textured in the (111) orientation (Fig.S5). The $V_g$-independent resistivity of B-Pt (Fig.S8) did not support a $V_g$-modulated spin Hall angle of the B-Pt as well [35].

We further measured the $V_g$-controlled MECD effect at different $T$. The $V_g$-tunability over the MECD effect was strongly depended on $T$ from 240 K to 300 K (Fig.4(a)). The difference in $\alpha$ under $V_g=\pm5$ V increased by a factor of 3.5 (from $0.6\times10^{-7}$ at 240 K to $2.1\times10^{-7}$ at 300 K) (Fig.4(b)). This strong $T$-dependence cannot favor the possibility of a $V_g$-controlled intrinsic spin Hall conductivity ($\sigma_{SH}^{int}$) since the electronic structure of Pt varies little with $T$. Nevertheless, the strong $T$-dependence can be naturally obtained as following. According to the spin-mixing conductance model across a NM/MI interface [4,6,26,27,36], the spin-torque-transfer efficiency and the spin-magnon convertance both depend on the $s$-$d$ exchange coupling strength and thus probability of electrons penetrating into the insulating YIG as evanescent states. The probability certainly depends on the interface barrier (thus $V_g$) and also $T$ since $T$ determines the kinetic energy of electrons in YIG. Supposing (1) $\pm5$ V gating leads to the similar band bending at different $T$ and (2) the classic thermal activation theory holds, we would expect an exponential $T$-dependence (Arrhenius law [37]) for the MECD coefficient. Fig.4(b) shows the fitting well matched the experimental data and the caused difference in the effective tunneling barrier by $\pm V_g$ reached 0.13 eV. Since the spin-mixing conductance depended on the $s$-$d$ coupling in the same way as the spin-magnon convertance, the SMR shared the same $V_g$-dependence as the MECD effect naturally. Band bending at interfaces relies on charged defect density which pins the Fermi level and influences bending degree, which probably accounts for the observation that a smoother and well-crystallized B-Pt/YIG interface (evidenced by a sharper electron diffraction pattern at this region) contributed to the $V_g$-controllability.

Now the above experimental data persuade us to attribute the $V_g$-controlled MECD effect to the $V_g$-induced changes in the spin-magnon conductance across the B-Pt/YIG interface. However, the measured $V_g$-$\alpha$ deviated from the theoretical prediction by the saturation trend at large $V_g$. We attribute this deviation to the redistributed voltage on the contact resistance ($R_{contact}$) since $I_{leak}$ increases divergently with $V_g$. In practice, we rewrote the Hamiltonian in YIG $H_{MI} = p^2/2m + V_0 + \Gamma \mathbf{S} \cdot \boldsymbol{\sigma} - ez\frac{V_g - I_{leak} \cdot R_{contact}}{t}$, considering the voltage dropped on $R_{contact}$. The calculated $G_r$-$V_g$ relation (Fig.4(c)) using parameters for Pt/YIG: Fermi energy $\varepsilon = 5$ eV, $V_0 = 5.5$ eV, $\Gamma=0.5$ eV [4] and $R_{contact} = 15$ MΩ agrees well with experiment. $G_r$ increased (decreased) with positive (negative) $V_g$ and saturated at $V_g \approx \pm 2\,\text{V}$. The calculated $G_r$ change by $V_g$ saturated at 13%/(MV/cm), also in a quantitative agreement with the experiment value ~10%/(MV/cm). The calculated result can also be well fitted with the hyperbolic tangent function with $|b/a|=0.041$ and $c=-0.45$ V$^{-1}$, which was the reason why we had used the hyperbolic tangent fitting to mathematically trace the $V_g$-dependences of $\alpha$, $V_{SP}^{peak}$ and SMR ratio.

In summary we have experimentally demonstrated a field-effect magnon transistor based on the MECD effect in the Pt/YIG/Pt sandwich. With the voltage-induced band bending of YIG, the energy profile of the B-Pt/YIG interfacial barrier and consequently its spin-magnon convertance was modulated. In this sense, the MECD effect was directly modulated by the gate voltage. Our finding promises direct modulation of spin-magnon conversion by electric fields, which shows a feasible pathway toward electrically controllable magnonics.

**Acknowledgements**: This work was financial supported by the National Key Research and Development Program of China [MOST Grant No. 2022YFA1402800], the National Natural Science Foundation of China [NSFC, Grant No. 51831012, 12134017], and partially supported by the Strategic Priority Research Program (B) of Chinese Academy of Sciences [CAS Grant No. XDB33000000, Youth Innovation Promotion Association of CAS (2020008)].

**Contributions**: X.F.H. led and was involved in all aspects of the project. Y.Z.W., J. D. and P. C. deposited stacks and fabricated devices. Y.Z.W. and C.H.W. conducted magnetic and transport property measurement. T.Y.Z., C.H.W. and Y.Z.W. contributed to modelling and theoretical analysis. C.H.W., Y.Z.W., T. Y. Z., G. Q. Y. and X.F.H. wrote the paper. X.F.H. and C.H.W. supervised and designed the experiments. All the authors contributed to data mining and analysis.

**Conflict of Interests**: The authors declare no competing interests.

**Methods**:

Sample Preparation:

The Si/SiO$_2$//Pt(10)/Y$_3$Fe$_5$O$_{12}$ (YIG)(80)/Pt(5 nm) heterostructures are deposited by ultrahigh vacuum magnetron sputtering system (ULVAC-MPS-400-HC7) with a base pressure < 5×10$^{-6}$ Pa. The bottom Pt Hall bar (B-Pt) with dimensions of 20×200 μm$^2$ was first fabricated on substrates by standard photolithography, followed by deposition of 80 nm YIG film. After deposition, a high-temperature annealing was carried out in an oxygen atmosphere to improve the crystalline quality of both YIG and Pt/YIG interface. Finally, another round of deposition and photolithography was carried out to fabricate top Pt Hall bar (T-Pt) with same dimensions. The terminal of T-Pt and B-Pt Hall bars are designed away from each other allow two Hall bars being input and detected independently. For the spin pumping device, B-Pt and T-Pt are fabricated into two independent stripes with dimensions of 10×360 μm$^2$ by the above mentioned method. And an 80 nm Au co-planar wave guide (CPW) was deposited afterwards. The two Pt strips are placed in the gap of the CPW.

Measurement of Magnetic Property:

The M-H hysteresis was measured with a vibrating sample magnetometer (VSM, MicroSense EZ-9) with field applied parallel to the film plane (IP curve) or perpendicular to film plane (OOP curve).

Measurement of Transport Property:

All the magnon mediate current drag (MCD) and spin magnetoresistance (SMR) test were carried out in a physical property measurement system (PPMS-9 T, Quantum design) with magnetic field up to 9 T and temperatures down to 1.8 K. During measurements, the input current was supplied by a Keithley 2400 source-meter while a Keithley 2182 nanovoltmeter detected the corresponding voltage. The gate voltage was provided by another Keithley 2400 across the Hall channel of T-Pt and B-Pt (grounded) Hall bars. The magnetic field was fixed at 1 T and sample rotated in xoy, xoz or yoz plane.

For angular dependent of MCD signal measurements, the input current ($I_{in}$) was applied in the long axis of B-Pt Hall bar and the voltage signal ($\Delta V$) was picked up alone the long axis of T-Pt Hall bar. Then the magnitude of MCD voltage $V_{drag}$ under certain $I_{in}$ was obtained by fitting the $\Delta V - \gamma$ curves measured at different $I_{in}$ with $\Delta V = V_{drag} \cos 2\gamma$. The magnon drag parameter was then calculated by $\alpha \equiv \frac{V_{drag}}{I_{in} R_{T-Pt}}$, where $R_{T-Pt}$ is the resistance of T-Pt electrode.

And for SMR test the resistance of the B-Pt (T-Pt) electrode was measured by four-terminal method, and the angular dependence of change in $R_{Pt}$ ($\Delta R_{Pt}$-$\gamma$) was well fitted by $\Delta R_{Pt} = R_{SMR}\cos 2\gamma$ and the SMR ratio was thus obtained by $|R_{SMR}/R_{Pt}|$.

The spin pumping test was carried out at room temperature in a home-build electromagnet with magnetic up to ~ 3500 Oe. A signal generator (ROHDE&SCHWARZ SMB 100A) supplies a microwave signal modulated with a 1.172 kHz signal to CPW and the voltage signal was picked up by a lock-in amplifier (Stanford SR830), while external magnetic field *H* applied perpendicular to the direction that spin pumping voltage was picked up. To minimize interference, $V_g$ was provided by dry batteries during spin pumping measurements.

**Data availability**: The data that support the findings of this study are available from the corresponding author upon reasonable request.